\documentclass[aip,reprint,nofootinbib]{revtex4-1}
\usepackage{natbib}%
\usepackage{graphicx}%
\usepackage{dcolumn}% Align table columns on decimal point
\usepackage{bm}% bold math
\begin{document}

%\preprint{}

\title{Photon spin-to-orbital angular momentum conversion via an electrically tunable $q-$plate} %Title of paper



\author{Bruno Piccirillo}
\email[]{bruno.piccirillo@na.infn.it}
%\thanks{}
\affiliation{Dipartimento di Scienze Fisiche, Universit\`{a} di Napoli "Federico II",\\
Complesso Universitario di Monte S. Angelo, 80126 Napoli, Italy}
\affiliation{CNISM-Consorzio Nazionale Interuniversitario per le Scienze Fisiche della Materia, Napoli}
\author{Vincenzo D'Ambrosio}
%\email[]{}
%\thanks{}
\affiliation{Dipartimento di Scienze Fisiche, Universit\`{a} di Napoli "Federico II",\\
Complesso Universitario di Monte S. Angelo, 80126 Napoli, Italy}
\author{Sergei Slussarenko}
%\email[]{}
%\thanks{}
\affiliation{Dipartimento di Scienze Fisiche, Universit\`{a} di Napoli "Federico II",\\
Complesso Universitario di Monte S. Angelo, 80126 Napoli, Italy}
\author{Lorenzo Marrucci}
%\email[]{}
%\thanks{}
\affiliation{Dipartimento di Scienze Fisiche, Universit\`{a} di Napoli "Federico II",\\
Complesso Universitario di Monte S. Angelo, 80126 Napoli, Italy}
\affiliation{CNR-SPIN Complesso Universitario di Monte S. Angelo, 80126 Napoli, Italy}
\author{Enrico Santamato}
%\email[]{}
%\thanks{}
\affiliation{Dipartimento di Scienze Fisiche, Universit\`{a} di Napoli "Federico II",\\
Complesso Universitario di Monte S. Angelo, 80126 Napoli, Italy}
\affiliation{CNISM-Consorzio Nazionale Interuniversitario per le Scienze Fisiche della Materia, Napoli}
\date{\today}

\begin{abstract}
Exploiting electro-optic effects in liquid crystals, we achieved real-time control of the retardation of liquid-crystal-based $q-$plates through an externally applied voltage. The newly conceived electro-optic $q$-plates can be operated as electrically driven converters of photon spin into orbital angular momentum, enabling a variation of the orbital angular momentum probabilities of the output photons over a time scale of milliseconds.
\end{abstract}

%\pacs{42.50Ar, 03.67Mn}% insert suggested PACS numbers in braces on next line

%\keywords{Spin angular momentum, orbital angular momentum, electro-optic effects, liquid crystals}%Use %showkeys class option if keyword

\maketitle %\maketitle must follow title, authors, abstract and \pacs

%\section{}
%\label{}
%\subsection{}
%\subsubsection{}

For years the orbital angular momentum (OAM) of photons has been consigned to a back seat in the study of both classical and quantum optics. At the very beginning, in combination with the spin angular momentum (SAM), it was adopted as an essentially auxiliary concept describing the properties of wave functions under rotations~\cite{landau4,jackson}. This was partly due to the fact that, in general, the spin and orbital contributions cannot be considered separately~\cite{jackson}, except within the paraxial approximation. However, even in the small-angle limit, the OAM, due to the dearth of tools suitable for its manipulation, has been used less than SAM, the latter being much handier than the former thanks to the availability of a harvest of polarization optical components. No doubt, the interest for both fundamental and applicative aspects of OAM has kept up with the development of methods and devices for its manipulation. At present, the importance of OAM for quantum optics is mainly due to the fact that it is defined on an infinite-dimensional Hilbert space and, by its own nature, could be suitable for implementing single-photon qudits. The use of high dimensional qudits instead of qubits is desirable since it may, for example, lead to simplifying quantum computations~\cite{muthukrishnan00,lanyon08} and improve quantum cryptography~\cite{bergman08}. The most commonly used methods for generating and manipulating the OAM are based on computer generated holograms (CGHs). However, in order to manage the OAM of light as much handily as possible, it has been recently introduced a liquid-crystal-based birefringent plate with retardation $\delta$ and optical axis unevenly oriented according to a distribution with topological charge $q$, after which the name $q$-plate (QP)~\cite{marrucci06}. A QP modifies the angular momentum state of an incident photon by giving to each of its circularly polarized components a finite probability, depending on the retardation $\delta$, of finding the photon in a polarization state with opposite helicity and OAM quantum number $l$ increased or decreased by the amount $\Delta l= 2q$, whether the initial helicity is positive or negative respectively. In $q$-plates with $q=1$, the optical axis distribution is cylindrically symmetric around the central defect and the total angular momentum of the incident photons is therefore conserved (SAM-to-OAM conversion or STOC). The adoption of such device in quantum optics has recently enabled the observation of two-photon Hong-Ou-Mandel coalescence interference of photons carrying non-zero OAM and the demonstration of the $1\rightarrow 2$ universal optimal quantum cloning of OAM-encoded qubits and qudits~\cite{nagali09a,nagali09b,nagali10}. Besides the topological charge $q$, a key-feature of a QP is the birefringent retardation, since it enables to regulate the probability of switching between $l$ and $l \pm 2 q$, i.e. the STOC efficiency, according to the wavelength of the input photons. High efficiencies for generation, manipulation and detection of OAM states are often desirable, especially when only few photons are available.

\begin{figure}
 \includegraphics[scale=0.9]{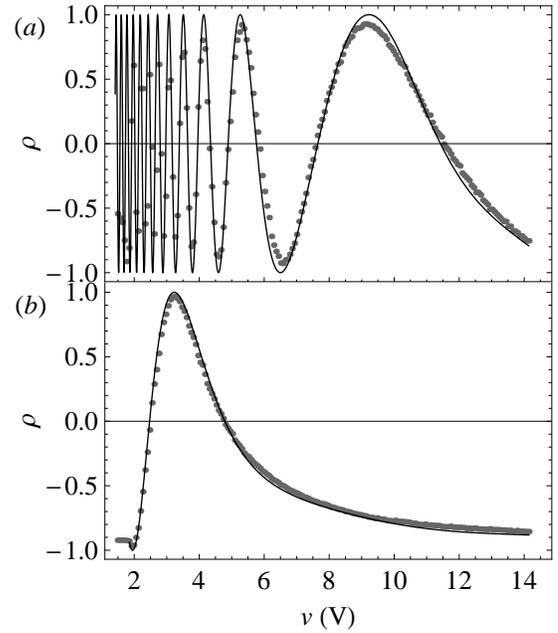}
 \caption{\label{fig:contrast} Measured contrast ratio $\rho(v) = \cos{\delta(v)}$, reported as function of $v$ for EOQP$_1$ (\textit{a}) and EOQP$_2$ (\textit{b}). The continuous lines represent the theoretical behaviors.}
\end{figure}
In this paper, to achieve a full control of the STOC efficiency, we present an electro-optical $q$-plate (EOQP) whose retardation may be changed through an externally applied voltage. Hitherto, STOC efficiencies exceeding $95\%$ (significantly higher than the efficiencies of the CGHs) were obtained tuning the $q$-plate retardation $\delta$ by controlling the material temperature~\cite{karimi09a}. A thermally-tuned QP has been successfully used with a Dove prism inserted into a Sagnac polarizing interferometer in order to generate arbitrary linear combinations of OAM eigenstates with $l = \pm 2$ by manipulating the polarization state of an input linearly polarized TEM$_{00}$ laser beam~\cite{karimi10a}. The thermal tunability of $\delta$ arises from the temperature dependence of the liquid crystal order parameter, and ultimately of the intrinsic birefringence of such material. The thermal control of a QP assures an easy-to-made stable retardation at the cost of a very slow time response. This is a limitation whenever the experiment requires a real-time variation of STOC efficiency, for instance, for qudit manipulation or when more wavelengths are involved. Electric-field-based regulation of $\delta$ enables one to overcome the time-response limitations imposed by the thermal method and make QPs suitable for more demanding tasks. The working principle of EOQP is based on the well-known property of external static electric or magnetic fields to change the orientation of the liquid-crystal molecular director $\bm n$, representing the local average orientation of liquid crystal molecules~\cite{degennes}. We fabricated and tested two EOQPs with $q=1$, so that the OAM impressed to converted photons is $l= \pm 2$. These devices have different thickness and have been manufactured by different methods. The first was a nominal 20~$\mu$m thick film of E7 liquid crystal from Merck Ltd, sandwiched between two ITO-coated glass substrates, beforehand coated with a polyimide for planar alignment and circularly rubbed, as described elsewhere~\cite{karimi09a}. The second EOQP was a nominal 6~$\mu$m thick film of E7, sandwiched between two ITO-coated glass substrates, beforehand coated with a polyimide UV-photo-aligned for planar topologically-charged optical axis distribution. The ITO transparent conductive films work as electrodes for the application of an electric field to the liquid crystal. From now on, the former device will be referred to as EOQP$_1$ and the latter as EOQP$_2$. Adopting essentially the same apparatus as that described in Ref.~\onlinecite{karimi09a}, for both the EOQPs, we measured, as a function of the applied voltage, the powers of the converted ($P_c$) and unconverted ($P_u$) components of the output beam for an incident $\lambda=532$~nm circularly polarized TEM$_{00}$ laser beam.
\begin{figure}
 \includegraphics[scale=0.87]{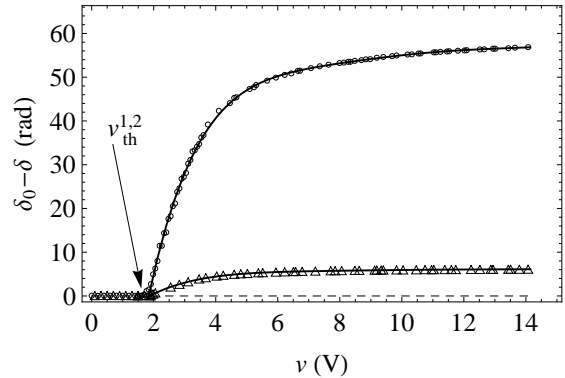}
 \caption{\label{fig:deltaV} The optical phase change $\delta (v)$ vs the applied voltage as extracted from the contrast $\rho(v)$ for EOQP$_1$ ($\circ$) and EOQP$_2$ ($\triangle$). The continuous lines correspond to sixth order polynomial curve obtained fitting the experimental data.}
\end{figure}
The temperature of both the EOQPs was maintained stable at 30$^\circ$~C during measurements. $P_u$ and $P_c$ are expected to depend on the optical retardation $\delta$ according to the Malus-like laws~\cite{karimi09a}:
\begin{equation}\label{eq:malus}
P_u = P_0 \cos^2(\delta/2), \hspace{1 cm} P_c = P_0 \sin^2(\delta/2),
\end{equation}
where $P_0$ is the total output power. In Fig.~\ref{fig:contrast}, the contrast ratios $\rho(v)=(P_u - P_c)/(P_u + P_c) = \cos{\delta(v)}$ are shown, for the rubbing-aligned and UV-photo-aligned EOQPs respectively, as functions of the rms values $v$ of the applied 1~kHz AC voltage~\cite{Note1}.
%\footnote{An AC rather than a DC voltage is applied to avoid electrochemical %degradation~\cite{blinov}.}
The dependence of $\delta(v)$ on the voltage arises from the torque exerted on the liquid crystal molecules by the applied electric field~\cite{degennes}. The behavior of $\delta(v)$ versus $v$, as deduced from $\rho(v)$, is shown in Fig.~\ref{fig:deltaV} for both devices. For $v < v_{th} = 1.85 \pm 0.01$~V, the nematic is undistorted in both the EOQPs and $\delta(v) = \delta_0 = 2\pi \Delta n L / \lambda + \delta_0^*$, where $L$ stands for the thickness of the cell, $\Delta n = 0.23$ is the intrinsic birefringence of the nematic E7 at 30$^\circ$~C and $\delta_0^*$ is a small constant residual birefringence. For EOQP$_1$, $\delta_0 \approx 1.7 \pi$ and for EOQP$_2$ $\delta_0 \approx 0.8 \pi$. For $v=v_{th}$ there is a discontinuity in the slope ${\rm d}\delta/{\rm d} v$. This is peculiar of a second order phase transition between the unperturbed and the distorted conformations at the critical voltage $v_{th}$ (Fre\'edericksz transition)~\cite{freedericksz33}. The theory of electro-optical effects in liquid crystals~\cite{degennes,blinov} predicts $v_{th}=\pi\sqrt{k_{11}/\epsilon_0 \epsilon_a}$,  where $k_{11}=9.2 \times 10^{-12}$~N (30$^\circ$~C) is the splay elastic constant of the liquid crystal~\cite{hakemi96}, $\epsilon_0$ is the vacuum dielectric constant and $\epsilon_a = 14.3$ at 1~kHz at 30$^\circ$~C~\cite{liu10}. Actually, such expression returns $v_{th} \lesssim 1$~V, i.e. half the experimental value. Such disagreement could be ascribed to the underlying simplified assumption of pure splay deformation. However, consistently with the theory, the experimental values of $v_{th}$, within the experimental uncertainties, turn out to be the same for both EOQPs, which differ from one another in the thickness only. In the limit of high applied voltage, the overall change $\Delta = \delta_0 - \delta(v) = \delta_0 - \delta_0^*=2\pi \Delta n L / \lambda$ is $\Delta \approx 56.8$ for EOQP$_1$ and $\Delta \approx 6.1$ for EOQP$_2$, yielding $L \approx 21$~$\mu$m for EOQP$_1$ and $L \approx 2.3$~$\mu$m for EOQP$_2$, respectively.

 The different thickness of the cells is responsible for the different saturation values of $\delta(v)$ and for their different switching-off times. Assuming, for simplicity, that the initial director alignment is homogeneous, the switching-off reorientation, in the small distortion limit, is expected to decay exponentially in time with a constant $\tau=\frac{\gamma_1}{k_{11}}\left(\frac{L}{\pi}\right)^2$, where $\gamma_1=150$~mPa (30$^\circ$~C) is the rotational viscosity of the liquid crystal~\cite{ran09}.
\begin{figure}
 \includegraphics[scale=0.9]{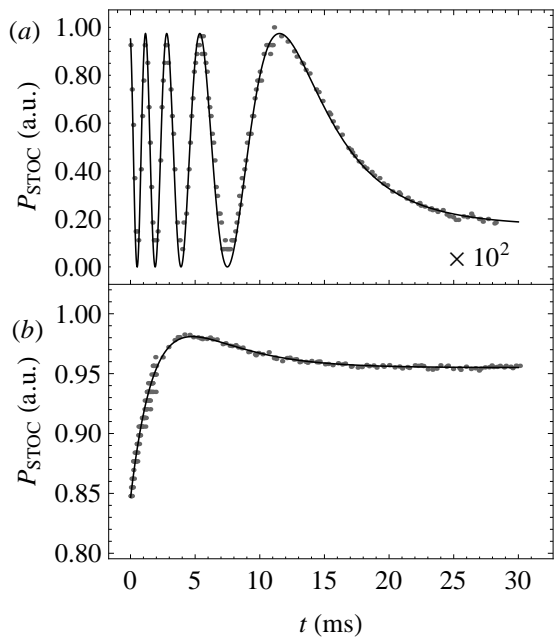}
 \caption{\label{fig:decay} Time decay of the power of the converted components of the output beam for EOQP$_1$ (\textit{a}) and EOQP$_2$ (\textit{b}) after the voltage switch-off. The starting voltages were $v_1=2.8$~V for EOQP$_1$ and $v_2=2.2$~V for EOQP$_2$. The continuous lines represent the theoretical behaviors. The time scale in panel (\textit{a}) is slower by a factor 100 with respect to panel (\textit{b}).}
\end{figure}
The time behavior of the switching-off $P_c$ signals is consistent with such prediction and the measured decay time constants are $\tau_1= 0.931 \pm 0.002$~s and $\tau_2 = (0.80 \pm 0.01){\times}10^{-2}$~s respectively (Fig.~\ref{fig:decay}). These values are in reasonable agreement with those expected from theory, i.e. $\tau_1^{theor}=0.53$~s and $\tau_2^{theor}=0.62 \times 10^{-2}$~s.
\begin{figure}
 \includegraphics[scale=0.87]{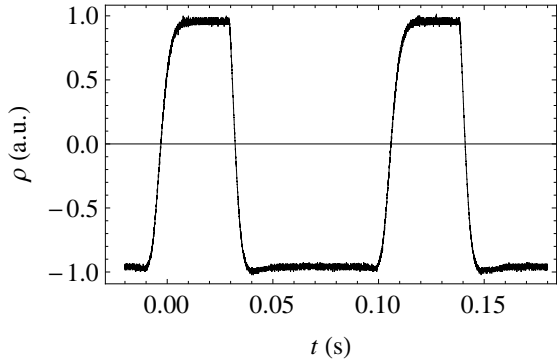}
 \caption{\label{fig:OnOff} Time behavior of contrast ratio $\rho$ for on-off switching of the applied voltage ($v_{on}=3.2$~V) for EOQP$_2$.}
\end{figure}
Nevertheless, consistently with the theoretical prediction for $\tau$, EOQP$_2$, due to its reduced thickness, switches off much faster than EOQP$_1$ and may be used for fast switching between different values of $\delta$ as shown in Fig.~\ref{fig:OnOff}. Finally, in order to check that the EOQPs actually work as SAM-to-OAM converters, the OAM of the output beam was directly measured adopting a method of sorting between $l=0$ and $l=\pm 2$ based on a polarizing Sagnac interferometer, with a Dove prism inserted along one of its arm, as elsewhere suggested~\cite{karimi10a}.
\begin{figure}
 \includegraphics[scale=0.87]{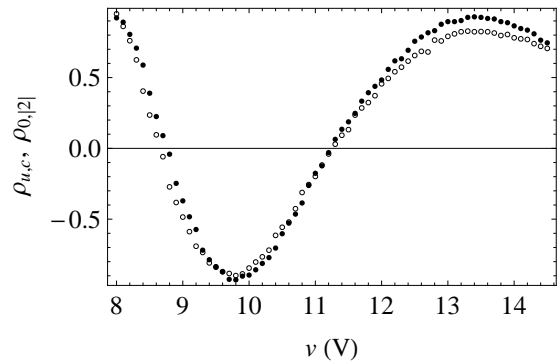}
 \caption{\label{fig:correlation} Contrast ratio $\rho_{u,c}(v)$ vs voltage for unconverted-converted components of the output sorted with respect to polarization ($\bullet$). Contrast ratio $\rho_{0,|2|}(v)$ vs voltage for OAM $l=0$ and $l=\pm 2$ components of the output beam measured with Dove-Sagnac sorter ($\circ$). Data refer to EOQP$_1$.}
\end{figure}
In Fig.~\ref{fig:correlation}, for EOQP$_1$, the measured contrast ratio between the $l=|2|$ and $l=0$ components of the output beam was reported, for comparison, together with the contrast ratio between the converted and unconverted components respectively, as sorted with respect to their polarizations as above mentioned. The correlation between the two signal is 99.6~$\%$.

In summary we realized an electro-optical SAM-to-OAM converter, aimed not only at improving the handiness of classical QPs, but also at performing new tasks exploiting the newly introduced capability of EOQPs of managing superpositions of OAM eigenstates with $l=0$, $l=2$ and $l=-2$ in real-time for high-dimensional qudits manipulation.\\

 We acknowledge the financial support of the FET-Open programme within the 7$^{th}$ Framework Programme of the European Commission, under grant number: 255914, Phorbitech.

% If in two-column mode, this environment will change to single-column format so that long equations can be displayed.
%\begin{widetext}
%$$\mbox{put long equation here}$$
%\end{widetext}

% Figures should be put into the text as floats.
% Use the graphics or graphicx packages (distributed with LaTeX2e).
% Put the label that you will use with \ref{} command in the braces of the \label{} command.
%
% \begin{figure}
% \end{figure}

% \begin{table}
% \end{tabular}
% \end{table}

%\begin{acknowledgments}
%\end{acknowledgments}

%\bibliographystyle{aip}
%\bibliography{jshort,general,OAM,OQ,lcbib}

\end{document}